\documentclass{iopart}
\usepackage{euscript,iopams,amssymb,amsfonts,graphicx,bm}
\usepackage{pgfplots,setstack}

\usepackage{float}
\usepackage{epsfig}
\usepackage{esint}
 \usepackage{tikz-cd}

\usepackage{bm,braket}

\eqnobysec

\newcommand{\dis}{\displaystyle}

\newcommand{\calP}{{\mathcal P}}

\newcommand{\calJ}{{\mathcal J}}
\newcommand{\R}{{\mathbb R}}

\newcommand{\X}{\mathbf{X}}
\renewcommand{\P}{\mathbb{P}}

\newcommand{\PP}{\widetilde{P}}

\newcommand{\x}{\mathbf{x}}
\newcommand{\y}{\mathbf{y}}
\renewcommand{\e}{{\mathrm e}}
\newcommand{\E}{{\mathbb E}}

\renewcommand{\P}{\mathbb P}
\newcommand{\p}{\widetilde{p}}

\renewcommand{\j}{\widetilde{j}}

\newcommand{\ellh}{\hat{\ell}}

\begin{document}

 \title[Encounter-based model of a run-and-tumble particle]{Encounter-based model of a run-and-tumble particle}

\author{Paul C. Bressloff}
\address{Department of Mathematics, University of Utah 155 South 1400 East, Salt Lake City, UT 84112}

\begin{abstract} 
In this paper we extend the encounter-based model of diffusion-mediated surface absorption to the case of an unbiased run-and-tumble particle (RTP) confined to a finite interval $[0,L]$ and switching between two constant velocity states $\pm v$ at a rate $\alpha$. The encounter-based formalism is motivated by the observation that various surface-based reactions are better modeled in terms of a reactivity that is a function of the amount of time that a particle spends in a neighborhood of an absorbing surface, which is specified by a functional known as the boundary local time. The effects of surface reactions are taken into account by identifying the first passage time (FPT) for absorption with the event that the local time crosses some random threshold $\widehat{\ell}$. In the case of a Brownian particle, the local time $\ell(t)$ is a continuous non-decreasing function of the time $t$. Taking $\Psi(\ell)\equiv \P[\widehat{\ell}>\ell]$ to be an exponential distribution, $\Psi[\ell]=\e^{-\kappa_0\ell}$, is equivalent to imposing a Robin boundary condition with a constant rate of absorption $\kappa_0$. One major difference in the encounter-based model of an RTP is that the boundary local time $\ell(t)$ is a now a discrete random variable that counts the number of collisions of the RTP with the boundary. Given this modification, we show that in the case of a geometric distribution $\Psi(\ell)=z^{\ell}$, $z=1/(1+\kappa_0/v)$, we recover the RTP analog of the Robin boundary condition. This allows us to solve the boundary value problem (BVP) for the joint probability density for particle position and the local time, and thus incorporate more general models of absorption based on non-geometric distributions $\Psi(\ell)$. We illustrate the theory by calculating the mean FPT (MFPT) for absorption at $x=L$ given a totally reflecting boundary at $x=0$. We also determine the splitting probability for absorption at $x=L$ when the boundary at $x=0$ is totally absorbing.

\end{abstract}

\maketitle
\section{Introduction}

A number of processes in cell biology can be modeled in terms of a particle switching between a left-moving and a right-moving state.
Examples include the growth and shrinkage of microtubules \cite{Dogterom93} or cytonemes \cite{Bressloff19}, the bidirectional motion of
molecular motors \cite{Newby10}, and the `run-and-tumble' motion of bacteria such as {\em E. coli} \cite{Berg04}. The simplest version of this class of active matter is the so-called unbiased run-and-tumble particle (RTP), in which the left-moving and right-moving states have the same speed $v$ and are equally likely. Studies at the single particle level include properties of the position density of a free RTP \cite{Martens12,Gradenigo19,Singh19}, non-Boltzmann stationary
states for an RTP in a confining potential \cite{Dhar19,Sevilla19,Dor19}, RTPs under stochastic resetting \cite{Evans18,Bressloff20,Santra20a} and the analysis of first-passage times (FPTs) \cite{Angelani14,Angelani15,Angelani17,Malakar18,Demaerel18,Doussal19}.
The FPT or stopping time of a general stochastic process is a random variable that determines when the stochastic process is terminated or killed. Examples of a stopping condition include a chemical or electrical signal crossing some threshold, a diffusing particle reaching a target boundary, and a particle escaping from a bounded domain through a small aperture \cite{Redner01,Bressloff13}. In the case of the FPT to reach a target boundary, one can implement the stopping condition by taking the boundary to be totally absorbing. However, in many real-world applications, the boundary is not perfectly
absorbing, which means that a particle can reach a given site on the boundary many times before reacting with it.  Hence, the stopping condition of the FPT is determined by the time the particle first reacts with the boundary rather than the time it first reaches the boundary. In other words, the boundary is now partially absorbing or reflecting, which is implemented using some form of radiation boundary condition. In the case of a diffusing particle, the latter is often referred to as a Robin boundary condition, in which the flux into the boundary is proportional to the density at the boundary, with the constant of proportionality identified as the reaction rate. On the other-hand, the analysis of non-diffusive stochastic processes with partially absorbing boundaries is much less developed.

In this paper we consider a first passage time (FPT) problem for unbiased RTP motion in a finite interval with a partially absorbing boundary at one end. The case of a constant rate of absorption has been considered previously \cite{Angelani17}, and is the analog of the Robin boundary condition for single-particle diffusion. Here we consider a much more general class of partially absorbing boundaries by extending the so-called encounter-based model of diffusion-mediated surface absorption \cite{Grebenkov19b,Grebenkov20,Grebenkov22,Bressloff22,Bressloff22b}. The latter is motivated by the observation that various surface-based reactions are better modeled in terms of a reactivity that is a function of the amount of time that a particle spends in a neighborhood of an absorbing surface, which is specified by a functional known as the boundary local time \cite{McKean75,Majumdar05}. The encounter-based model of diffusion-mediated absorption considers the joint probability density or generalized propagator $P(\x,\ell,t)$ for the pair $(\X(t),\ell(t))$, in the case of a perfectly reflecting boundary $\partial \Omega$, where $\X(t)\in \Omega \subset \R^d$ and $\ell(t)\in [0,\infty)$ denote the particle position and local time, respectively. The latter is defined as
\begin{equation}
\label{locD}
\fl \ell(t)=\lim_{h\rightarrow 0} \frac{1}{h} \int_0^tH(h-\mbox{dist}(\X(\tau),\partial \Omega)d\tau=\int_0^t \left \{ \int_{\partial \Omega}\delta(\X(\tau)-\y)d\y \right \}d\tau.
\end{equation} 
where $H(x)$ is the Heaviside function and $\delta(\x)$ is the multidimensional Dirac delta function. It can be shown that $\ell(t)$ exists and is a continuous, non-decreasing  function of time \cite{Ito65}.
The effects of surface reactions are then incorporated 
 by introducing the stopping time 
${\mathcal T}=\inf\{t>0:\ \ell(t) >\widehat{\ell}\}$,
 with $\widehat{\ell}$ a random local time threshold. Given the probability distribution $\Psi(\ell) = \P[\ellh>\ell]$ with $\Psi(0)=1$, the marginal probability density for particle position is defined according to
 \begin{equation}
 \label{pc}
 p(\x,t)=\int_0^{\infty} \Psi(\ell)P(\x,\ell,t)d\ell.
 \end{equation}
 Laplace transforming the propagator with respect to $\ell$, which is equivalent to taking an exponential distribution $\Psi(\ell)=\e^{-u\ell}$, leads to a classical Robin boundary value problem (BVP) for the corresponding diffusion equation, in which the reactive component has a constant rate of absorption $\kappa_0=u$. Hence, one can incorporate more general forms of absorption by solving the propagator BVP for a constant absorption rate $\kappa_0$, setting $\kappa_0=u$, inverting the Laplace transform with respect to $u$, and then calculating the marginal probability density for a general distribution $\Psi(\ell)$. 
 
In this paper we show how all of the steps of the encounter-based approach have natural analogs when $\X(t)$ represents the position of an RTP rather than a Brownian particle. The one major difference is that the local time $\ell(t)$ of an RTP is a discrete rather than a continuous random variable \cite{Singh21}. This means that the integral on the right-hand side of equation (\ref{pc}) is replaced by a discrete sum (under the rescaling $\ell \rightarrow v\ell$):
 \begin{equation}
 \label{pd}
 p(\x,t)=\sum_{\ell =0}^{\infty}\Psi(\ell-1)P(\x,\ell,t),
 \end{equation} 
 with $\Psi(-1)=1$. In addition, we find that a constant rate of absorption $\kappa_0$, which was previously analyzed in \cite{Angelani17}, corresponds to taking the distribution of the local time threshold to a be a geometric distribution, that is, $\Psi(\ell)=z^{\ell}$, with $z=1/(1+\kappa_0/v)$.

The structure of the paper is as follows. In section 2 we consider the classical FPT problem for an RTP in a finite interval $x\in [0,L]$ with a totally reflecting boundary at $x=0$ and a partially absorbing boundary with a constant absorption rate at $x=L$. In section 3 we develop the encounter-based model of RTP absorption in terms of the local time propagator. In particular, we derive the propagator BVP, and show that in the case of a geometric distribution $\Psi(\ell)$ we recover the RTP analog of the Robin boundary condition analyzed in section 2. This allows us to solve the propagator BVP and thus incorporate more general models of absorption based on non-geometric distributions $\Psi(\ell)$. We illustrate the theory by calculating the mean FPT (MFPT) $\tau$ for absorption at $x=L$ given a totally reflecting boundary at $x=0$ (section 4). In particular, we find that $\tau=\tau_{\infty}+2\E[\ell]L/v$, where $\tau_{\infty}$ is the corresponding MFPT for a totally absorbing boundary and $\E[\ell]=\sum_{\ell=0}^{\infty}\ell\psi(\ell)$, $\psi(\ell)=\Psi(\ell-1)-\Psi(\ell)$. Hence, the MFPT only exists if $\psi(\ell)$ has a finite first moment. We also determine the splitting probability for absorption at $x=L$ when the boundary at $x=0$ is totally absorbing (section 5).

\setcounter{equation}{0}
\section{Run-and-tumble particle in an interval with a single absorbing boundary}

Consider a particle confined to the interval $x\in [0,L]$ that randomly switches between two constant velocity state labeled by $\sigma=\pm$ with $v_+=v$ and $v_-=-v$ for some $v>0$. Furthermore, suppose that the particle reverses direction according to a Poisson process with rate $\alpha$. The position $X(t)$ of the particle at time $t$ then evolves according to the piecewise deterministic equation
\begin{equation}
\label{PDMP}
\frac{dX}{dt}=vn(t),
 \end{equation}
where $n(t)=\pm 1$ is a dichotomous noise process that switches sign at the rate $\alpha$. Following other authors, we will refer to a particle whose position evolves according to equation (\ref{PDMP}) as a run-and-tumble particle (RTP). Let $p_{\sigma}(x,t|x_0)$ be the probability density of the RTP at position $x\in [0,L]$ at time $t>0$ and moving to the right ($\sigma=+)$ and to the left ($\sigma =-$), respectively. The associated differential Chapman-Kolomogorov (CK) equation is then
\numparts
\begin{eqnarray}
\label{DLa}
\frac{\partial p_{+}}{\partial t}&=-v \frac{\partial p_{+}}{\partial x}-\alpha p_{+}+\alpha p_{-},\\
\frac{\partial p_{-}}{\partial t}&=v \frac{\partial p_{-}}{\partial x}-\alpha p_{-}+\alpha p_{+}.
\label{DLb}
\end{eqnarray}
\endnumparts
This is supplemented by the initial conditions $x(0)=x_0$ and $\sigma(0)=\pm $ with probability $\rho_{\pm }=1/2$. Note that under the change of variables
\begin{equation}
\label{cov}
\fl p(x,t|x_0)=p_+(x,t|x_0)+p_-(x,t|x_0),\quad j(x,t|x_0)=v[p_+(x,t|x_0)-p_-(x,t|x_0)],
\end{equation}
with $p$ the marginal probability for particle position and $j$ the probability flux, we can rewrite equations (\ref{DLa}) and (\ref{DLb}) in the form
\numparts
\begin{eqnarray}
\label{JDLa}
\frac{\partial p}{\partial t}&=-\frac{\partial j}{\partial x},\\
\frac{\partial j}{\partial t}&=-v^2 \frac{\partial p}{\partial x}-2\alpha j.
\label{JDLb}
\end{eqnarray}

It remains to specify the boundary conditions at the ends $x=0,L$. We will focus on the case of a single partially absorbing boundary at $x=L$ and a totally reflecting boundary at $x=0$. (In section 5, we will modify the latter by considering a totally absorbing boundary at $x=0$.) We thus have
\begin{equation}
\label{JBC0}
 j(0,t|x_0)=0 ,
 \end{equation}
and
\begin{equation}
\label{JBC}
 j(L,t|x_0) =\kappa_0 p_-(L,t|x_0)=\frac{\kappa_0}{2} \left (p(L,t|x_0)-\frac{j(L,t|x_0)}{v}\right ),
\end{equation}
\endnumparts
where $\kappa_0$ is a constant reaction rate.
In the limit $\kappa_0\rightarrow \infty$, the boundary at $x=L$ becomes totally absorbing with $p_-(L,t)=0$, whereas it is totally reflecting when $\kappa_0=0$. Note that partial absorption has been defined along analogous lines to the classical Robin boundary condition for the diffusion equation. From a microscopic perspective, the totally reflecting boundary condition at $x=0$ means that whenever the particle hits the boundary its velocity is immediately reversed. On the other hand, at the partially reflected boundary $x=L$, the particle is reflected for the first $n$ times it hits the boundary and is absorbed at the $(n+1)$-th collision, where $n$ is a random variable. This probabilistic interpretation will be expanded upon and generalized in section 3. Here we wish to calculate the MFPT to be absorbed at $x=L$ having started at $x_0\in (0,L)$.  The MFPT can be determined either by solving a backward CK equation for the survival probability or by Laplace transforming the forward CK equation. We will follow the latter approach here. (Note that the FPT problem for an RTP with partially absorbing boundaries at both ends was previously solved by Angelani \cite{Angelani15}. For completeness, we include the details of our analysis here, since they are needed for the more general encounter-based approach developed in subsequent sections.)

Laplace transforming equations (\ref{JDLa}) and (\ref{JDLb}) with $\p(x,s)=\int_0^{\infty}\e^{-st}p(x,t)dt$ etc. gives
\begin{eqnarray}
\label{jp}
  \frac{\partial \j}{\partial x} =-s \p + \delta(x-x_0), \quad
D\frac{\partial \p}{\partial x}= -\j  ,
\end{eqnarray}
where  
\begin{equation}
\label{D}
D=D(s) \equiv \frac{v^2}{2\alpha +s}
\end{equation}
is an effective diffusivity that depends on the Laplace variable $s$. The Laplace transformed boundary condition at $x=L$ is
\begin{eqnarray}
\label{jpb}
\left [2+\frac{\kappa_0}{v}\right ] \j(L,s|x_0)=\kappa_0\p(L,s|x_0).
\end{eqnarray}
It follows that $\p$ satisfies the second-order equation
\begin{equation}
\label{diff}
D\frac{\partial^2 \p}{\partial x^2}-s\p=-\delta(x-x_0) .
\end{equation}
This has the general solution
\begin{equation}
\label{ps}
\p(x,s|x_0)=A_n\cosh(\sqrt{s/D} x)+G_n(x,s|x_0),
\end{equation}
where $G_n(x,s|x_0)$ is a Green's function of the modified Helmholtz equation on $[0,L]$:
\begin{eqnarray}
\label{GG}
\fl D\frac{\partial^2 G_n(x,s|x_0)}{\partial x^2}-sG_n=-\delta(x-x_0),\ \partial_xG_n(0,s|x_0)=0,\ G_n(L,s|x_0)=0.
\end{eqnarray}
We have also imposed the no-flux boundary condition at $x=0$.
The constant $A_n$ is determined by the boundary condition (\ref{jpb}).
In addition, $\p(L,s|x_0)=A_n\cosh(\sqrt{s/D} L)$ and
\begin{eqnarray}
\label{js}
\fl \j(L,s|x_0)&=-\left . D\frac{\partial \p(x,s|x_0)}{\partial x}\right |_{x=L}
=-A_n\sqrt{sD}\sinh(\sqrt{s/D} L)-D\partial_xG_n(L,s|x_0).
\end{eqnarray}
Hence, after some algebra we find that $A_n=A_n(s)$ with
\begin{eqnarray}
\label{As}
A_n(s)\equiv -\frac{\dis D\partial_xG_n(L,s|x_0)\left (2+\frac{\kappa_0}{v} \right )}{\dis   \sqrt{sD}\left (2+\frac{\kappa_0}{v}\right )\sinh(\sqrt{s/D} L) +\kappa_0 \cosh(\sqrt{s/D} L)}.
\end{eqnarray}

The next step is to introduce the survival probability
\begin{equation}
\label{SP}
S(x_0,t)=\int_0^{L}p(x,t|x_0)dx.
\end{equation}
If $T(x_0)$ denotes the first passage time to be absorbed starting at $x_0$, then ${\rm Prob}[T(x_0)\leq t] =1-S(x_0,t)$. The FPT density is thus given by
\begin{equation*}
f(x_0,t)=-\frac{\partial S(x_0,t)}{\partial t}=\int_0^{L}\frac{\partial p(x,t|x_0)}{\partial t}dx= j(L,t|x_0),
\end{equation*}
and the corresponding MFPT is
\begin{eqnarray}
\fl \tau(x_0):= \E[ T(x_0)] &= \int_0^{\infty}t f(x_0,t)dt= -\int_0^{\infty}t\frac{\partial S(x_0,t)}{\partial t}dt =\int_0^{\infty} S(x_0,t)dt,
\end{eqnarray}
after integration by parts. Hence,
\begin{eqnarray}
 \tau(x_0)=\int_0^{L} \p(x,0|x_0)dx.\end{eqnarray}
Taking the limit $s\rightarrow 0$ in equation (\ref{ps}) with $A=A(s)$ given by equation (\ref{As}) yields
\begin{eqnarray}
\p(x,0|x_0)= -\frac{\dis \Lambda_n(x_0)\left (2+\frac{\kappa_0}{v}\right )}{\dis  \kappa_0 }+G_n(x,0|x_0),
\end{eqnarray}
with $\Lambda_n(x_0)\equiv \lim_{s\rightarrow 0}D\partial_xG_n(L,s|x_0)$. Integrating both sides of the first equation in (\ref{GG}) with respect to $x$ implies that
\begin{equation}
D\partial_xG_n(L,s|x_0)=s\int_0^LG_n(x,s|x_0)dx-1.
\end{equation}
That is, $\Lambda_n(x_0)=-1$ since $\lim_{s\rightarrow 0}sG_n(x,s|x_0)=0$. Finally, $G_n(x,0|x_0)$ has the explicit solution
\begin{equation}
\fl G_n(x,0|x_0)=\frac{1}{D_0}\left \{(L-x_0)H(x_0-x)+(L-x)H(x-x_0)\right \},\quad D_0=\frac{v^2}{2\alpha},
\end{equation} 
where $H(x)$ is the Heaviside function. Combining our various results, we obtain the following formula for the MFPT:
\begin{equation}
\label{tau0}
\tau(x_0)=\frac{2L}{\kappa_0} +\tau_{\infty}(x_0),\quad \tau_{\infty}(x_0)=\frac{L}{v}+\frac{L^2-x_0^2}{2D_0},
\end{equation}
where $\tau_{\infty}(x_0)$ is the MFPT for the RTP with a totally absorbing boundary at $x=L$. The expression for $\tau(x_0)$ is a special case of the more general formula derived in \cite{Angelani15}.

It is well-known that in the double limit $v^2\rightarrow \infty$ and $\alpha \rightarrow \infty$ with $D_0$ fixed, equations (\ref{JDLa})--(\ref{JBC}) reduce to the diffusion equation with a Robin boundary condition at $x=L$:
\numparts
\begin{eqnarray}
\label{RBa}
&\frac{\partial p}{\partial t}=D_0\frac{\partial^2p}{\partial x^2},\quad j(x,t|x_0)=-D_0\frac{\partial p(x,t|x_0)}{\partial x},\\
&j(0,t|x_0)=0,\quad j(L,t|x_0) = \frac{\kappa_0}{2} p(L,t|x_0).
\label{RBb}
\end{eqnarray}
\endnumparts
Taking the diffusion limit of equation (\ref{tau0}) shows that the corresponding MFPT is still given by the first equation in (\ref{tau0}) except that now $ \tau_{\infty}(x_0)=(L^2-x_0^2)/(2D_0)$.

   \section{Encounter-based formulation of RTP absorption}

Recently, it has been shown how to reformulate the Robin boundary condition for a diffusing particle using a probabilistic interpretation based on the so-called boundary local time \cite{Grebenkov19b,Grebenkov20,Grebenkov22,Bressloff22,Bressloff22a}. The latter is a Brownian functional that keeps track of the amount of time a particle spends in a local neighborhood of a boundary \cite{McKean75,Majumdar05}. This encounter-based method essentially reformulates partially reflected Brownian motion in terms of totally reflected Brownian motion combined with a local time absorption mechanism. The latter is independent of the dynamics in the bulk.In this section we develop a corresponding reformulation for an RTP in $[0,L]$ with a partially absorbing boundary at $x=L$. 

For the moment, suppose that the boundary at $x=L$ to be totally reflecting. Given the position $X(t)$ of the particle at time $t$, we define the scaled local time at the boundary $x=L$ according to (see also \cite{Singh21})
\begin{equation}
\label{loc}
\ell(t) =v \int_0^t\delta(X(\tau)-L)d\tau .
\end{equation}
Each trajectory of the RTP will typically yield a different value of $\ell(t)$, which means that $\ell(t)$ is itself a stochastic process. In contrast to the local time (\ref{locD}) of Brownian motion, $\ell(t)$ is a discrete random variable that is equal to the number of times the RTP hits the boundary $x=L$ in the time interval $[0,t]$, see also Ref. \cite{Singh21}.
Introduce the joint probability density or local time propagator
\begin{eqnarray}
\fl P_{\sigma}(x,\ell,t|x_0)dx =\P[x<X(t)<x+dx,\ell(t)=\ell,\sigma(t)=\sigma|X(0)=x_0,\ell(0)=0].
\end{eqnarray}
Since the local time only changes at the boundary, the evolution equation within the bulk of the domain takes the same form as for $p_{\sigma}(x,t|x_0)$. In terms of the transformed propagators
\numparts
\begin{eqnarray}
P(x,\ell,t|x_0)&=P_+(x,\ell,t|x_0)+P_-(x,\ell,t|x_0),\\ J(x,\ell,t|x_0)&=v[P_+(x,\ell,t|x_0)-P_-(x,\ell,t|x_0)],
\end{eqnarray}
\endnumparts
we have
\numparts
 \begin{eqnarray} 
   \label{JPCK1}
   \frac{\partial P}{\partial t}   &=&- \frac{\partial J}{\partial x},\\
 \frac{\partial J}{\partial t}   &=&-v^2 \frac{\partial P}{\partial x}-2\alpha J .
   \label{JPCK2}
  \end{eqnarray}
  \endnumparts
  Moreover, the reflecting boundary condition at $x=0$ is
  \begin{equation}
  \label{BC0}
  J(0,\ell,t|x_0)=0.
  \end{equation}
  
  \begin{figure}[b!]
  \raggedleft
   \includegraphics[width=8cm]{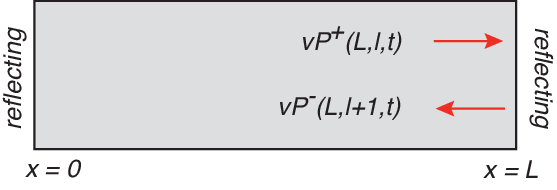}
  \caption{Schematic diagram showing the fluxes entering and leaving the boundary at $x=L$. (The interval is represented as a rectangular domain for illustrative purposes.)}
  \label{fig1}
  \end{figure}

The nontrivial step is determining the boundary condition at $x=L$. In the case of an RTP, this can be derived by considering the balance of probability fluxes at the boundary, see Fig. \ref{fig1}. The flux hitting the boundary at time $t$ and with the local time $\ell$, $\ell \geq 0$, is $vP_+(L,\ell,t)$. This is immediately converted to a leftward flux $vP_-(L,\ell+1,t)$ with the local time $\ell+1$ due to the collision with the totally reflecting boundary. We thus have the balance condition
\begin{eqnarray}
 0&=v[P_+(L,\ell,t|x_0)-P_-(L,\ell+1,t|x_0)] \nonumber \\
&=v[P_+(L,\ell,t|x_0)-P_-(L,\ell,t|x_0)]\nonumber \\
 &\qquad +v[P_-(L,\ell,t|x_0)-P_-(L,\ell+1,t|x_0)].
\end{eqnarray}
Rearranging this equation gives
\begin{eqnarray}
\label{BCL}
 J(L,\ell,t|x_0) 
&=v[P_-(L,\ell+1,t|x_0)-P_-(L,\ell,t|x_0)]. 
\end{eqnarray}
We also have the condition $P_-(L,0,t|x_0)=0$ for $x_0\neq L$, since the probability of a left-moving particle being found at $x=L$ is zero until there has been at least one collision with the boundary ($\ell>0$).
It is important to note the difference between the totally and partially reflecting boundary conditions (\ref{BC0}) and (\ref{BCL}). This is due to the fact that we are only keeping track of the local time accumulated at $x=L$, since this is needed in order to incorporate partial absorption at $x=L$. Indeed, if we were to sum equation (\ref{BCL}) with respect to $\ell$ then we would lose any information about $\ell$ and recover the standard reflecting boundary condition $J(L,t|x_0)=\sum_{\ell=0}^{\infty} J(L,\ell,t|x_0)=0$.

 We now introduce the discrete Laplace transforms 
\begin{equation}
\label{0LT}
\fl  \widetilde{P}(x,z,t|x_0)=\sum_{\ell=0}^{\infty} z^{\ell} P(x,\ell,t|x_0),\quad  \widetilde{J}(x,z,t|x_0)=\sum_{\ell=0}^{\infty} z^{\ell}  J(x,\ell,t|x_0),
\end{equation}
with $z\in [0,1]$, and similarly for $\widetilde{P}_{\pm}(x,z,t|x_0)$. The transformed propagator BVP is
\numparts
 \begin{eqnarray} 
   \label{JPCK1LT}
 \frac{\partial \widetilde{P}}{\partial t}   &=&- \frac{\partial \widetilde{J}}{\partial x},\\
 \frac{\partial \widetilde{J}}{\partial t}   &=&-v^2 \frac{\partial \widetilde{P}}{\partial x}-2\alpha \widetilde{J},
   \label{JPCK2LT}
  \end{eqnarray}
  with
   \begin{equation}
   \label{BCLLT}
\widetilde{J}(0,z,t|x_0)=0,\quad \widetilde{J}(L,z,t|x_0)=\frac{v[1-z]}{z}\widetilde{P}_-(L,z,t|x_0).
\end{equation}
 \endnumparts
 The boundary condition at $x=L$ is obtained as follows:
 \begin{eqnarray*}
\fl &\sum_{\ell=0}^{\infty} z^{\ell}  [P_-(L,\ell+1,t|x_0)-P_-(L,\ell,t|x_0)]\\
\fl &=z^{-1}\sum_{\ell'=1}^{\infty} z^{\ell'}  P_-(L,\ell',t|x_0)-\widetilde{P}_-(L,z,t|x_0)=\frac{v[1-z]}{z}\widetilde{P}_-(L,z,t|x_0),
\end{eqnarray*}
since $P_-(L,0,t|x_0)=0$.
 Comparison with the BVP given by equations (\ref{JDLa})--(\ref{JBC}) shows that
if we set 
\begin{equation}
\frac{v[1-z]}{z}=\kappa_0 \Rightarrow z =z(\kappa_0)\equiv \frac{v}{v+\kappa_0},
\end{equation}
then the solution of the BVP for a RTP with a constant rate of absorption is related to the Laplace transformed generalized propagators according to 
\begin{equation}
\label{loco}
\fl p(x,t|x_0)= \widetilde{P}(x,z(\kappa_0),t|x_0),\quad j(x,t|x_0)= \widetilde{J}(x,z(\kappa_0),t|x_0) .
\end{equation}

Following along analogous lines to the encounter-based formulation of diffusion \cite{Grebenkov19b,Grebenkov20,Grebenkov22,Bressloff22}, we can give a probabilistic interpretation of the above result. First, introduce the stopping time 
\begin{equation}
\label{Tell0}
{\mathcal T}=\inf\{t>0:\ \ell(t) >\widehat{N}\},
\end{equation}
 with $\widehat{N}$ a random variable that represents a threshold for the number of boundary collisions. Let 
 \begin{equation}
 \psi(n)=\P[\widehat{N}=n],\quad \sum_{n=0}^{\infty}\psi(n)=1
 \end{equation}
 and set
 \begin{equation}
\Psi(m) =\P[\widehat{N}>m]=\sum_{n=m+1}^{\infty} \psi(n)=1-\sum_{n=0}^m\psi(n).
\end{equation}
It follows that
\begin{equation}
\psi(n)=\Psi(n-1)-\Psi(n).
\end{equation}
The stopping time ${\mathcal T}$ is a random variable that specifies the time of absorption. The joint probability density for particle position and the velocity state can be expressed as
\begin{equation}
\fl p_{\pm}(x,t|x_0)dx=\P[X(t) \in (x,x+dx), \ \sigma(t)=\pm ,\ t < {\mathcal T}|X_0=x_0].
\end{equation}
Given that $\ell(t)$ is a nondecreasing process, the condition $t < {\mathcal T}$ is equivalent to the condition $\ell(t)<\widehat{N}$. This implies that
\begin{eqnarray*}
\fl p_{\pm}(x,t|x_0)dx&=\P[X(t) \in (x,x+dx),\ \sigma(t)=\pm, \ \ell(t) \leq \widehat{N}|X_0=x_0]\\
\fl &=\sum_{n=0}^{\infty} \psi(n)\P[X_t \in (x,x+dx),\ \sigma_t=\pm , \ \ell(t) \leq n |X_0=x_0]\\
\fl &=\sum_{n=0}^{\infty} \psi(n)\sum_{m=0}^n  P_{\pm}(x,m,t|x_0)\, dx.
\end{eqnarray*}
Using the identity
\begin{equation}
\sum_{n=0}^{\infty}f(n)\sum_{m=0}^n   g(m)=\sum_{m=0}^{\infty}g(m)\sum_{n=m}^{\infty}  f(n) 
\label{fg}
\end{equation}
for arbitrary functions $f,g$, it follows that
\begin{equation}
\label{bob}
p_{\pm}(x,t|x_0)=\sum_{m=0}^{\infty}\Psi(m-1)P_{\pm}(x,m,t|x_0),
\end{equation}
with $\Psi(-1)=1$.
Equations (\ref{bob}) are equivalent to equations (\ref{loco}) if we take $\psi(\ell)$ to be the geometric distribution
\begin{equation}
\label{odd}
\psi(n)=(1-z)z^n \Rightarrow \Psi(m-1)=z^m, \, m\geq 1.
\end{equation}

In conclusion, the probability density $p_{\pm}(x,t|x_0)$ for a constant rate of absorption $\kappa_0$ can be expressed in terms of the discrete Laplace transform of the local time propagator $P_{\pm}(x,\ell,t|x_0)$ with respect to the discrete local time $\ell$, since the partially absorbing boundary condition (\ref{JBC}) maps to a geometrical law for the threshold local time $\widehat{N}$. 
The advantage of the probabilistic formulation of a partially absorbing boundary is that one can consider a more general probability distribution $\psi(n) $ such that the marginal density becomes
  \begin{equation}
  \label{Boo}
 \fl  p_{\pm }^{\Psi}(x,t|x_0)=\sum_{\ell=0}^{\infty} \Psi(\ell-1)P_{\pm}(x,\ell,t|x_0)=\sum_{\ell=0}^{\infty} \Psi(\ell-1){\mathcal L}_{\ell}^{-1}\widetilde{P}_{\pm}(x,z,t|x_0),
  \end{equation}
  where ${\mathcal L}^{-1}_{\ell}$ denotes the inverse discrete Laplace transform. Hence, we can incorporate a more general model of absorption by solving the corresponding propagator BVP given by equations (\ref{JPCK1LT})--(\ref{BCLLT}), inverting the discrete Laplace transforms, and then evaluating the sums with respect to $\ell$. Finally, note that given a discrete Laplace transform $\widetilde{f}(z)$, the inverse transform is defined by a contour integral on the unit circle $C$:
  \begin{equation}
  f(\ell)={\mathcal L}_{\ell}^{-1} [\widetilde{f}(z)]=\oint_C\frac{\widetilde{f}(z)}{2\pi i z^{\ell+1}}dz.
  \end{equation}
  For relatively simple transforms $\widetilde{f}(z)$, we can extract the inverse transform by expanding $\widetilde{f}(z)$ as a geometric series in $z$.

\paragraph{Continuum approximation for the local time.} One of the interesting differences between the encounter-based model of an RTP and the corresponding model of a Brownian particle is that in the latter case the local time is a continuous variable. This means that the marginal probability density $ p(x,t|x_0)$ for a constant rate of absorption $\kappa_0$ is obtained by Laplace transforming the propagator with respect to the continuous local time and identifying the Laplace variable with $\kappa_0$. The corresponding local time threshold is generated by an exponential distribution \cite{Grebenkov19b,Grebenkov20,Grebenkov22,Bressloff22,Bressloff22}. A closer connection between the two models can be made in the small-$\kappa_0$ regime. Since the probability that the RTP is absorbed following one collision event is very small,  it is likely that multiple collisions occur before absorption. In other words, we have $\P[\ell\gg 1]\approx 1$. Returning to the boundary condition (\ref{BCL}), we can treat $\ell$ as a continuous variable and take
\begin{eqnarray}
\label{BCL2}
 J(L,\ell,t|x_0) 
&=v\frac{\partial P_-(L,\ell,t|x_0)}{\partial \ell}. 
\end{eqnarray}
Introducing the Laplace transforms
\begin{equation}
\label{c0LT}
\fl  \widetilde{P}(x,u,t|x_0)=\int_0^{\infty} \e^{-u \ell} P(x,\ell,t|x_0)d\ell,\quad  \widetilde{J}(x,u,t|x_0)=\int_0^{\infty} \e^{-u \ell} J(x,\ell,t|x_0)d\ell
\end{equation}
etc., the boundary condition at $x=L$ becomes
\begin{equation}
   \label{BCLLT2}
\widetilde{J}(L,u,t|x_0)=vu\widetilde{P}_-(L,u,t|x_0),
\end{equation}
which implies that $u=\kappa_0/v$. Comparison with the exact analysis using discrete Laplace transforms implies that $\e^{-\kappa_0/v}\approx 1/(1+\kappa_0/v)$, which is valid provided that $\kappa_0\ll v$. 

 \setcounter{equation}{0}
\section{Calculation of the MFPT for generalized absorption} 

It turns out that the simplest 
way to proceed is to perform a double Laplace transform with respect to $t$ and $\ell$ by setting
\begin{eqnarray}
\label{dLP}
\calP(x,z,s|x_0)&=\int_0^{\infty} dt\, \e^{-st}\sum_{\ell=0}^{\infty} z^{\ell} P(x,\ell,t|x_0),\\
\calJ(x,z,s|x_0)&=\int_0^{\infty}dt\, \e^{-st}\sum_{\ell=0}^{\infty} z^{\ell}\calJ(x,\ell,t|x_0).
\label{dLJ}
\end{eqnarray}
This yields the propagator BVP
\begin{eqnarray} 
   \label{calPCK}
\frac{\partial\calJ}{\partial x}=  -s\calP +\delta(x-x_0) ,\quad D\frac{\partial \calP}{\partial x}= -\calJ,
  \end{eqnarray}
  with $D=D(s)$ given by equation (\ref{D}) and
  \begin{eqnarray}
  \label{caljs}
  \calJ(0,z,s|x_0)=0,\quad \calJ(L,z,s|x_0)=\frac{v(1-z)}{1+z}\calP(L,z,s|x_0),
  \end{eqnarray}
  which is formally identical to the BVP given by equations (\ref{jp}) and (\ref{jpb}), after setting $z=z(\kappa_0)$.
 It immediately follows from equations (\ref{ps}), (\ref{js}) and (\ref{As}) that the solution of the propagator BVP is
\begin{equation}
\label{calps}
\calP(x,z,s|x_0)=A(z,s)\cosh(\sqrt{s/D} x)+G_n(x,s|x_0),
\end{equation}
where $G_n(x,s|x_0)$ is the Green's function defined in equation (\ref{GG}),
\begin{eqnarray}
A(z,s)&\equiv -\frac{\dis D(1+z)\partial_xG_n(L,s|x_0)}{\dis   (1+z)\sqrt{sD}\sinh(\sqrt{s/D} L) +v(1-z)\cosh(\sqrt{s/D} L)}\nonumber \\
&=-\frac{\dis D(1+z)\partial_xG_n(L,s|x_0)}{\dis v\cosh(\sqrt{s/D} L)+  \sqrt{sD}\sinh(\sqrt{s/D} L)}\frac{1}{1-\Gamma(s)z},
\label{calAs}
\end{eqnarray}
and
\begin{equation}
\fl \Gamma(s)=\frac{v\cosh(\sqrt{s/D} L)-\sqrt{sD}\sinh(\sqrt{s/D} L)}{v\cosh(\sqrt{s/D} L)+\sqrt{sD}\sinh(\sqrt{s/D} L) }=\frac{v-\sqrt{sD}\tanh(\sqrt{s/D} L)}{v+\sqrt{sD}\tanh(\sqrt{s/D} L)}.
\end{equation}
Note that 
\begin{equation}
\label{Garm}
\Gamma(0)=1,\quad \Gamma'(0)=-\frac{2L}{v}.
\end{equation}

The survival probability that the particle hasn't been absorbed in the time interval $[0,t]$, having started at $x_0$, is defined according to
\begin{equation}
\label{S1}
S(x_0,t)=\sum_{\ell=0}^{\infty} \Psi(\ell)\left [\int_0^LP(x,\ell,t|x_0)dx\right ].
\end{equation}
Differentiating both sides of this equation with respect to $t$ implies that
\begin{eqnarray}
\fl \frac{\partial S(x_0,t)}{\partial t}&=\sum_{\ell=0}^{\infty}\Psi(\ell-1)\left [\int_0^L\frac{\partial P(x,\ell,t|x_0)}{\partial t}dx\right ]\nonumber \\
\fl &=-\sum_{\ell=0}^{\infty} \Psi(\ell-1)\left [\int_0^L\frac{\partial J(x,\ell,t|x_0)}{\partial x}dx\right ] =-\sum_{\ell=0}^{\infty} \Psi(\ell-1) J(L,\ell,t|x_0) \nonumber \\
\fl &=-v\sum_{\ell=0}^{\infty} \Psi(\ell-1) [P_-(L,\ell+1,t|x_0)-P_-(L,\ell,t|x_0)] \nonumber \\
\fl &=-v\sum_{\ell=1}^{\infty} [\Psi(\ell-2)-\Psi(\ell-1)] P_-(L,\ell,t|x_0)\nonumber \\
\fl & =-v\sum_{\ell=1}^{\infty}\psi(\ell-1)  P_-(L,\ell,t|x_0)\equiv - j^{\psi}(L,t|x_0).
\label{Q2}
\end{eqnarray}
The last line follows from equation (\ref{BCL}).
Laplace transforming equation (\ref{Q2}) with respect to $t$ and noting that $S(x_0,0)=1$ gives
\begin{equation}
\label{QL}
s\widetilde{S}(x_0,s)-1=-\widetilde{j}^{\psi}(L,s|x_0).
\end{equation}
The MFPT for absorption is then 
\begin{eqnarray}
\label{MFPT1}
\tau(x_0)&= \widetilde{S}(x_0,0)=\lim_{s\rightarrow 0}\frac{1-\widetilde{j}^{\psi}(L,s|x_0)}{s}=-\left .\frac{\partial}{\partial s}\widetilde{j}^{\psi}(L,s|x_0)\right |_{s=0}
\end{eqnarray}
with
\begin{eqnarray}
\fl \widetilde{j}^{\psi}(L,s|x_0)&=v\sum_{\ell=1}^{\infty}\psi(\ell-1) {\mathcal L}_{\ell}^{-1} \calP_-(L,z,s|x_0)\nonumber \\
\fl &=\frac{v}{2}\sum_{\ell=1}^{\infty}\psi(\ell-1)  {\mathcal L}_{\ell}^{-1} \left [\calP(L,z,s|x_0)-v^{-1}\calJ(L,z,s|x_0)\right ].
\end{eqnarray}

 \begin{figure}[b!]
  \raggedleft
   \includegraphics[width=8cm]{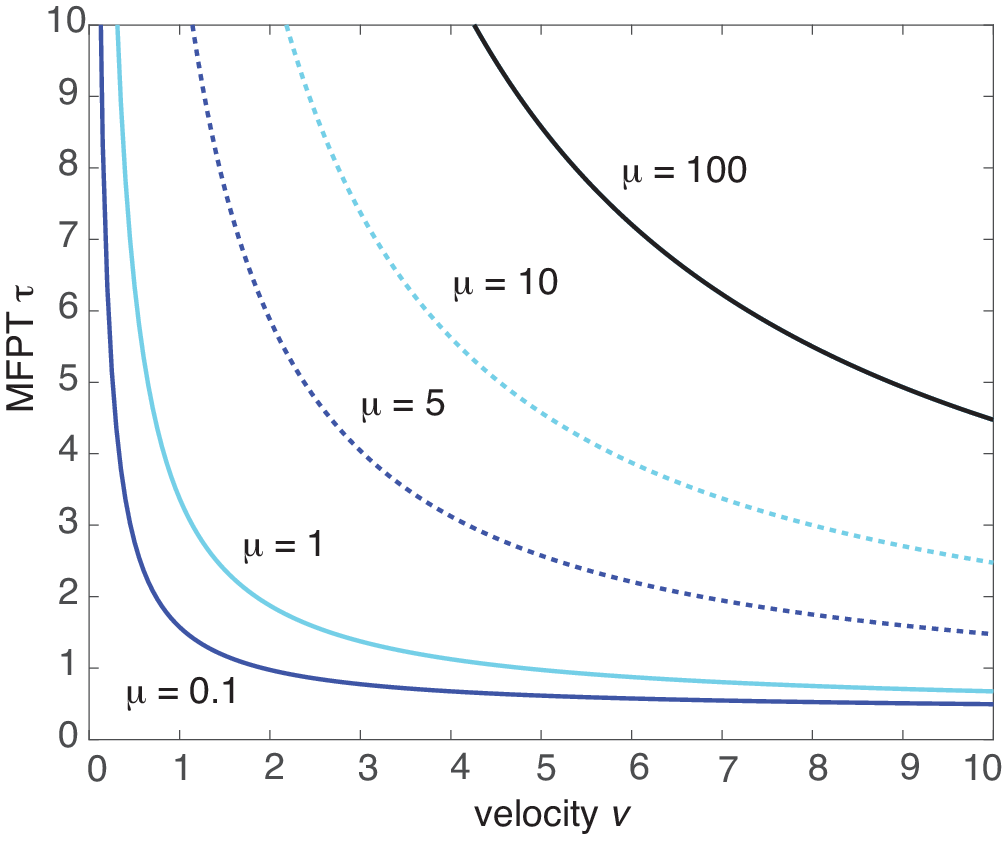}
  \caption{RTP in an interval with a partially absorbing boundary at $x=L$ and a totally reflecting boundary at $x=0$. Plot of the MFPT $\tau(x_0)$ as a function of the velocity $v$ for a distribution $\psi(\ell) $ having different means $\mu\equiv \E[\ell]$. Other parameters are $L=1$, $\alpha=1$ and $x_0=0.5$.}
  \label{fig2}
  \end{figure}
  
Substituting the solutions (\ref{calps}) and (\ref{caljs}) for $\calP(L,z,s|x_0)$ and $\calJ(L,z,s|x_0)$ with $G_n(L,s|x_0)=0$, and expanding $A(z,s)$ as a geometric series in $z$ gives
\begin{eqnarray}
\fl \widetilde{j}^{\psi}(L,s|x_0)&=-\frac{\dis vD\partial_xG_n(L,s|x_0)\cosh(\sqrt{s/D} L)}{\dis v\cosh(\sqrt{s/D} L)+  \sqrt{sD}\sinh(\sqrt{s/D} L)} \sum_{\ell=1}^{\infty}\psi(\ell-1)\Gamma(s)^{\ell-1} \nonumber \\
\fl &=- \frac{D\partial_xG_n(L,s|x_0)}{[1+(\sqrt{sD}/v)\tanh(\sqrt{s/D}L)]}\widetilde{\psi}(\Gamma(s)).
\end{eqnarray}
Differentiating with respect to $s$, taking the limit $s\rightarrow 0$ with $\lim_{s\rightarrow 0}D\partial_xG_n(L,s|x_0)=-1$, and substituting into equation (\ref{MFPT1}) yields
\begin{eqnarray}
\fl\tau(x_0)&= \widetilde{\psi}(1)\lim_{s\rightarrow 0}\bigg [D\partial_s\partial_xG_n(L,s|x_0)\nonumber \\
\fl &\quad +\frac{v^{-1}\sqrt{D/s}\tanh(\sqrt{s/D} L)+(L/v)\mbox{sech}^2(\sqrt{s/D}L)}{2[1+(\sqrt{sD}/v)\tanh(\sqrt{s/D}L)]^2}\bigg ]- \widetilde{\psi}'(1)\lim_{s\rightarrow 0} \Gamma'(s)\nonumber \\
\fl &=\widetilde{\psi}(1)\left [\frac{L^2-x_0^2}{2D_0}+\frac{L}{v}\right ]+\frac{2L}{v}\widetilde{\psi}'(1)\nonumber\\
\fl &=\frac{L^2-x_0^2}{2D_0} +\frac{L}{v}+\frac{2L}{v}\E[\ell] =\tau_{\infty}(x_0)+\frac{2L}{v}\E[\ell] .
\label{taugen}
\end{eqnarray}
We have used equations (\ref{Garm}) and the fact that $\widetilde{\psi}(z)$ is the moment generator of the local time. In particular,
\begin{equation}
\widetilde{\psi}(1)=\sum_{n=0}^{\infty} \psi(n)=1,\quad \widetilde{\psi}'(1)=\sum_{n=0}^{\infty} n\psi(n).
\end{equation}
Equation (\ref{taugen}) implies that a necessary condition for the existence of $\tau(x_0)$ is that $\psi(\ell)$ has a finite first moment, $\E[\ell]<\infty$. Clearly $\tau(x_0)\rightarrow \infty$ as $\E[\ell] \rightarrow \infty$ (totally reflecting boundary). On the other hand, $\tau(x_0)\rightarrow \tau_{\infty}(x_0)$ as $\E[\ell]\rightarrow 0$ (totally absorbing boundary).

We now fix the mean by setting $ \E[\ell]=\mu$ and express the MFPT as a function of $v$ with $D_0=v^2/2\alpha$:
  \begin{equation}
  \tau(v)=\alpha\frac{L^2-x_0^2}{v^2}+\frac{L}{v}+\frac{2\mu L}{v}.
\end{equation}
Example plots of the MFPT $\tau$ as a function of the velocity $v$ are shown in Fig. \ref{fig2} for different values of $\mu$. In the case of a geometric distribution, we have
\begin{equation}
\label{Geom}
\fl \psi(\ell)=(1-\lambda_0)\lambda_0^{\ell} \quad \Rightarrow \quad \widetilde{\psi}(z)=\frac{1-\lambda_0}{1-\lambda_0z}\quad \Rightarrow\quad  \E[\ell] =\frac{\lambda_0}{1-\lambda_0}
\end{equation}
Since the boundary conditions (\ref{JBC}) hold, we can relate the distribution parameter $\lambda_0$ to a constant absorption rate $\kappa_0$ according to $\lambda_0=1/(1+\kappa_0/v)$. Hence, substituting for $\E[\ell]$ in equation (\ref{taugen}) recovers equation (\ref{tau0}). The expression for the MFPT in equation (\ref{taugen}) still holds for any other discrete distribution $\psi(\ell)$ with a finite first moment. However, it is no longer possible to relate the distribution parameters to a constant absorption rate. A simple example of a non-geometric distribution with finite moments is the Poisson distribution:
\begin{equation}
\label{Poiss}
\psi(\ell)=\lambda_0^{\ell}\frac{\e^{-\lambda_0}}{\ell!} \quad \Rightarrow \quad \widetilde{\psi}(z)=\e^{\lambda_0(z-1)} \quad \Rightarrow\quad  \E[\ell] =\lambda_0.
\end{equation}

\section{Splitting probabilities for absorption at both ends}

Let us now consider an RTP with a totally absorbing boundary at $x=0$ and a partially absorbing boundary at $x=L$. In contrast to the BVP of section 2, the probability that the RTP is absorbed at $x=L$ is no longer unity. Therefore, it is necessary to determine the splitting probability that the particle is absorbed at $x=L$ before ever reaching (being absorbed at) $x=0$. 

\subsection{Constant rate of absorption at $x=L$}

Recall from section 3 that the double Laplace transform of the local time propagator satisfies a BVP that is identical in form to the BVP for the marginal densities $p(x,t|x_0)$ and $j(x,t|x_0)$ in the case of a constant rate of absorption $\kappa_0$ at $x=L$. After Laplace transforming the latter with respect to $t$, we have the BVP
\begin{equation}
\label{sdiff}
D\frac{\partial^2 \p}{\partial x^2}-s\p=-\delta(x-x_0),\quad \j =-D\frac{\partial \p}{\partial x},
\end{equation}
together with the boundary conditions
\begin{eqnarray}
\label{sjpb}
  \j(0,s|x_0)=-v\p(0,s|x_0),\quad \j(L,s|x_0)=\frac{\kappa_0}{2+\kappa_0/v}\p(L,s|x_0).
\end{eqnarray}
The boundary condition at $x=0$ is equivalent to $\p_+(0,s|x_0)=0$.
The general solution takes the form
 \begin{equation}
\label{psd}
\fl \p(x,s|x_0)=A_d\cosh(\sqrt{s/D} x)+B_d\sinh(\sqrt{s/D} x)+G_d(x,s|x_0),
\end{equation}
where 
\begin{eqnarray}
\label{GGd}
\fl D\frac{\partial^2 G_d(x,s|x_0)}{\partial x^2}-sG_d=-\delta(x-x_0),\ G_d(0,s|x_0)=0= G_d(L,s|x_0).
\end{eqnarray}
The Green's function $G_d$ has the explicit solution
\begin{equation}
\label{Gdsol}
\fl G_d(x,s|x_0)=\left \{ \begin{array}{cc} \frac{\displaystyle\sinh(\sqrt{s/D} x)\sinh(\sqrt{s/D} [L-x_0])}{\displaystyle\sqrt{sD}\sinh(\sqrt{s/D} L)} & \ x <x_0\\
\frac{\displaystyle\sinh(\sqrt{s/D} x_0)\sinh(\sqrt{s/D} [L-x])}{\displaystyle\sqrt{sD}\sinh(\sqrt{s/D} L)} & \ x >x_0 \end{array}\right . .
\end{equation}
The constants $A_d$ and $B_d$ are determined by the boundary conditions (\ref{sjpb}):
\numparts
\begin{eqnarray}
\label{sjs1}
vA_d=B_d\sqrt{sD}+D\partial_xG_d(0,s|x_0),
\end{eqnarray}
and
\begin{eqnarray}
\label{sjs2}
\fl&\kappa_0[ A_d\cosh(\sqrt{s/D} L)+B_d\sinh(\sqrt{s/D} L)]\\
\fl &=(2+\kappa_0/v)\left (-A_d\sqrt{sD}\sinh(\sqrt{s/D} L)-B_d\sqrt{sD}\cosh(\sqrt{s/D} L)-D\partial_xG_d(L,s|x_0)\right ).\nonumber 
\end{eqnarray}
\endnumparts
Substituting for $A_d$ in equation (\ref{sjs2}) gives
\begin{eqnarray*}
\fl &\kappa_0\left [v^{-1}[B_d\sqrt{sD}+D\partial_xG_d(0,s|x_0)]\cosh(\sqrt{s/D} L)+B_d\sinh(\sqrt{s/D} L)\right ]\\
\fl &=-(2+\kappa_0/v)\bigg([B_d\sqrt{sD}+D\partial_xG_d(0,s|x_0)]\frac{\sqrt{sD}}{v}\sinh(\sqrt{s/D} L)\\
\fl &\quad +B_d \sqrt{sD} \cosh(\sqrt{s/D} L)+D\partial_xG_d(L,s|x_0)\bigg ).
\end{eqnarray*}
Rearranging this equation shows that
\begin{eqnarray}
\fl B_d=-\frac{(2+\kappa_0/v)D\partial_xG_d(L,s|x_0)+ D\partial_xG_d(0,s|x_0)\Theta_3(s)}{(2+\kappa_0/v)\sqrt{sD}\Theta_1(s)+z_0\Theta_2(s)},
\end{eqnarray}
where
\numparts
\begin{eqnarray}
\Theta_1(s)&= \cosh(\sqrt{s/D} L)+\frac{\sqrt{sD}}{v}\sinh(\sqrt{s/D} L),\\ \Theta_2(s)&=\sinh(\sqrt{s/D} L)+\frac{\sqrt{sD}}{v}\cosh(\sqrt{s/D} L)\\
\Theta_3(s)&=(2+\kappa_0/v)\frac{\sqrt{sD}}{v}\sinh(\sqrt{s/D} L)+\frac{\kappa_0}{v}\cosh(\sqrt{s/D} L).
\end{eqnarray}
\endnumparts

Let  $\Pi_0(x_0,t)$ and $\Pi_L(x_0,t)$ denote, respectively, the probability that the particle is absorbed at $x=0$ and $x=L$ after time $t$, having started at $x_0$. Then
\begin{equation}
\label{PI}
\fl \Pi_{0}(x_0,t):=-\int_t^{\infty} j(0,t'|x_0)dt',\quad \Pi_{L}(x_0,t):=\int_t^{\infty} j(L,t'|x_0)dt' .
\end{equation}
In particular,  the splitting probabilities are
\begin{equation}
\fl\pi_{0}(x_0)=\Pi_{0}(x_0,0)=-\lim_{s\rightarrow 0}\j(0,s|x_0),\quad \pi_{L}(x_0)=\Pi_{L}(x_0,0)=\lim_{s\rightarrow 0}\j(L,s|x_0).
\end{equation}
Using equations (\ref{sjpb}) and (\ref{psd}) we find that
\begin{eqnarray}
\fl\pi_0(x_0)&=v\lim_{s\rightarrow 0}\p(0,s|x_0)=v\lim_{s\rightarrow 0}A_d\nonumber \\
\fl&=\lim_{s\rightarrow 0}[B_d\sqrt{sD}+D\partial_xG_d(0,s|x_0)] \\
\fl&=-\lim_{s\rightarrow 0}\frac{(2+\kappa_0/v)D\partial_xG_d(L,s|x_0)+(\kappa_0/v)D\partial_xG_d(0,s|x_0)}{2\kappa_0/v+\kappa_0L/D+2}\nonumber \\
\fl &\quad +\lim_{s\rightarrow 0}D\partial_xG_d(0,s|x_0).\nonumber
\end{eqnarray}
Finally, differentiating equation (\ref{Gdsol}) with respect to $x$ gives
\begin{eqnarray}
\lim_{s\rightarrow 0}D\partial_xG(0,s|x_0)&=\frac{L-x_0}{L},\quad \lim_{s\rightarrow 0}D\partial_xG(L,s|x_0)=-\frac{x_0}{L}.
\end{eqnarray}
Hence,
\begin{equation}
\label{pie}
\pi_0(x_0)=\frac{\displaystyle 2+\frac{\kappa_0}{v}+\frac{\kappa_0[L-x_0]}{D_0}}{\displaystyle 2+\frac{\kappa_0L}{D_0}+\frac{2\kappa_0}{v}}.
\end{equation}
It can be checked that $\pi_L(x_0)=1-\pi_0(x_0)$.
As expected, in the limit $\kappa_0\rightarrow 0$, we have $\pi_0(x_0)\rightarrow 1$ since the boundary at $x=L$ becomes totally reflecting. On the other hand, if $\kappa_0\rightarrow \infty$, then both boundaries are totally absorbing and
\begin{equation}
\pi_0(x_0)=\frac{1}{2}\frac{\displaystyle 1+\frac{v[L-x_0]}{D_0}}{\displaystyle 1+\frac{vL}{2D_0}}.
\end{equation}
Clearly, if the RTP starts at the center of the domain so that $x_0=L/2$ then $\pi_0(x_0)=1/2$ for all $v$ by symmetry.

 Finally, given the splitting probabilities, one can also construct a pair of conditional MFPTs for absorption at a specific end. It is then necessary that expectation is only taken with respect to trajectories that are never absorbed at the other end. The conditional MFPT for absorption at $x=L$ is thus defined according to
\begin{eqnarray}
T_L(x_0)&=-\frac{1}{\pi_L(x_0)}\int_0^{\infty} t\frac{\partial \Pi_L(x_0,t)}{\partial t}dt=\frac{1}{\pi_L(x_0)}\int_0^{\infty}    \Pi_L(x_0,t) dt\nonumber \\
&=\frac{1}{\pi_L(x_0)}\lim_{s\rightarrow 0}\widetilde{\Pi}_L(x_0,s),
\label{pT1}
\end{eqnarray}
after integrating by parts. Similarly, the conditional MFPT for absorption at $x=0$ is
\begin{eqnarray}
\label{pT0}
T_0(x_0)=\frac{1}{\pi_0(x_0)} \lim_{s\rightarrow 0}\widetilde{\Pi}_0(x_0,s).
\end{eqnarray}
Finally, differentiating equation (\ref{PI}) with respect to $t$ and then Laplace transforming with respect to $t$ implies that
\begin{equation}
\fl s\widetilde{\Pi}_0(x_0,s)-\pi_{0}(x_0)=\j(0,s|x_0),\quad s\widetilde{\Pi}_L(x_0,s)-\pi_{L}(x_0)=-\j(L,s|x_0).
\end{equation}
Rearranging and taking the limit $s\rightarrow 0$, we find
\numparts
\begin{eqnarray}
\label{piT0}
 \lim_{s\rightarrow 0}\widetilde{\Pi}_0(x_0,s)&=\left. \frac{\partial \j(0,s|x_0)}{\partial s}\right |_{s=0}=-v\left. \frac{\partial \p(0,s|x_0)}{\partial s}\right |_{s=0},\\ \lim_{s\rightarrow 0}\widetilde{\Pi}_L(x_0,s)&=-\kappa_0\left. \frac{\partial \p(L,s|x_0)}{\partial s}\right |_{s=0}.
 \label{piT1}
\end{eqnarray}
\endnumparts
The resulting expressions for the conditional MFPTs are considerably more cumbersome than the splitting probabilities, so we will focus on the latter.

\subsection{Generalized absorption at $x=L$} 
We now calculate the splitting probabilities when the partially absorbing boundary at $x=L$ is formulated in terms of the local time propagator. The doubly Laplace transformed propagator BVP is given by equations (\ref{calPCK}) with boundary conditions
  \begin{eqnarray}
\fl  \calJ(0,z,s|x_0)=-v\calP(0,z,s|x_0),\quad \calJ(L,z,s|x_0)=\frac{v(1-z)}{1+z}\calP(x,z,s|x_0),
  \end{eqnarray}
  Comparison of the propagator BVP with equations (\ref{sdiff}) and (\ref{sjpb})
implies that the splitting probability in the case of a general stopping local time distribution $\Psi(\ell)$ is
\begin{eqnarray}
\fl \pi_0(x_0)&=-\lim_{s\rightarrow 0} \sum_{\ell=0}^{\infty} \Psi(\ell-1){\mathcal L}_{\ell}^{-1} \calJ(0,z,s|x_0) =v\lim_{s\rightarrow 0}  \sum_{\ell=0}^{\infty} \Psi(\ell-1){\mathcal L}_{\ell}^{-1} \calP(0,z,s|x_0)\nonumber \\
\fl &=  \sum_{\ell=0}^{\infty}\Psi(\ell-1){\mathcal L}_{\ell}^{-1} \left (\frac{\displaystyle \frac{1+z}{z}+\frac{(1-z)}{z}\frac{v[L-x_0]}{D_0}}{\displaystyle \frac{[1+z]}{z}+\frac{(1-z)}{z}\left [\frac{vL}{D_0}+1\right ]}\right ) \\
\fl &=  \sum_{\ell=0}^{\infty}\Psi(\ell-1){\mathcal L}_{\ell}^{-1} \left (\frac{\displaystyle  1+ \frac{v[L-x_0]}{D_0}+z\left (1- \frac{v[L-x_0]}{D_0}\right ) }{\displaystyle  2+vL/D_0}\frac{1}{1-\Lambda_0 z}\right ),\nonumber
\end{eqnarray}
where
\begin{equation}
\Lambda_0=\frac{vL/D_0}{2+vL/D_0}.
\end{equation}
Performing the geometric series expansion of the last term we can invert the Laplace transform to obtain the result
\begin{eqnarray}
\fl \pi_0(x_0) &=\frac{1}{\displaystyle  2+vL/D_0}  \bigg\{ \left (   1+ \frac{v[L-x_0]}{D_0}\right ) \sum_{\ell=0}^{\infty}\Psi(\ell-1)\Lambda_0^{\ell}\nonumber \\
\fl &\hspace{2.5cm} +\left (1- \frac{v[L-x_0]}{D_0}\right  )\sum_{\ell=1}^{\infty}\Psi(\ell-1)\Lambda_0^{\ell-1} \bigg\},\nonumber\\
\fl &=\frac{1}{\displaystyle  2+vL/D_0} \left \{\frac{v[L-x_0]}{D_0}  \sum_{l=0}^{\infty}[\Psi(\ell-1)-\Psi(\ell)]\Lambda_0^{\ell}+\sum_{l=0}^{\infty}[\Psi(\ell-1)+\Psi(\ell)]\Lambda_0^{\ell} \right \}\nonumber \\
\fl &=\frac{1}{\displaystyle  2+vL/D_0} \left \{\frac{v[L-x_0]}{D_0} \sum_{l=0}^{\infty}\psi(\ell)\Lambda_0^{\ell}+\left (1+\frac{1}{\Lambda_0}\right )\sum_{l=0}^{\infty}\Psi(\ell-1)\Lambda_0^{\ell} -\frac{1}{\Lambda_0} \right \}\nonumber \\
\fl &=\frac{1}{\displaystyle  2+vL/D_0} \left \{\frac{v[L-x_0]}{D_0} \widetilde{\psi}(\Lambda_0)+\left (1+\frac{1}{\Lambda_0}\right )\widetilde{\Psi}(\Lambda_0) -\frac{1}{\Lambda_0} \right \}\nonumber \\
\fl &=1+\frac{1}{\displaystyle  2+vL/D_0} \left [\frac{v[L-x_0]}{D_0} -\frac{1+\Lambda_0}{1-\Lambda_0}\right ]\widetilde{\psi}(\Lambda_0).
\end{eqnarray}
We have used the identity
\begin{equation}
 \widetilde{\psi}(z)=\widetilde{\Psi}(z)(1-1/z)+1/z,
 \end{equation}
 which follows from Laplace transforming equation (\ref{odd}). Note that from the definition of $\Lambda_0$ we have
 \[\frac{1+\Lambda_0}{1-\Lambda_0}=   1+vL/D_0,\]
 so that $\pi_0(x_0)<1$.

 \begin{figure}[t!]
  \raggedleft
   \includegraphics[width=8cm]{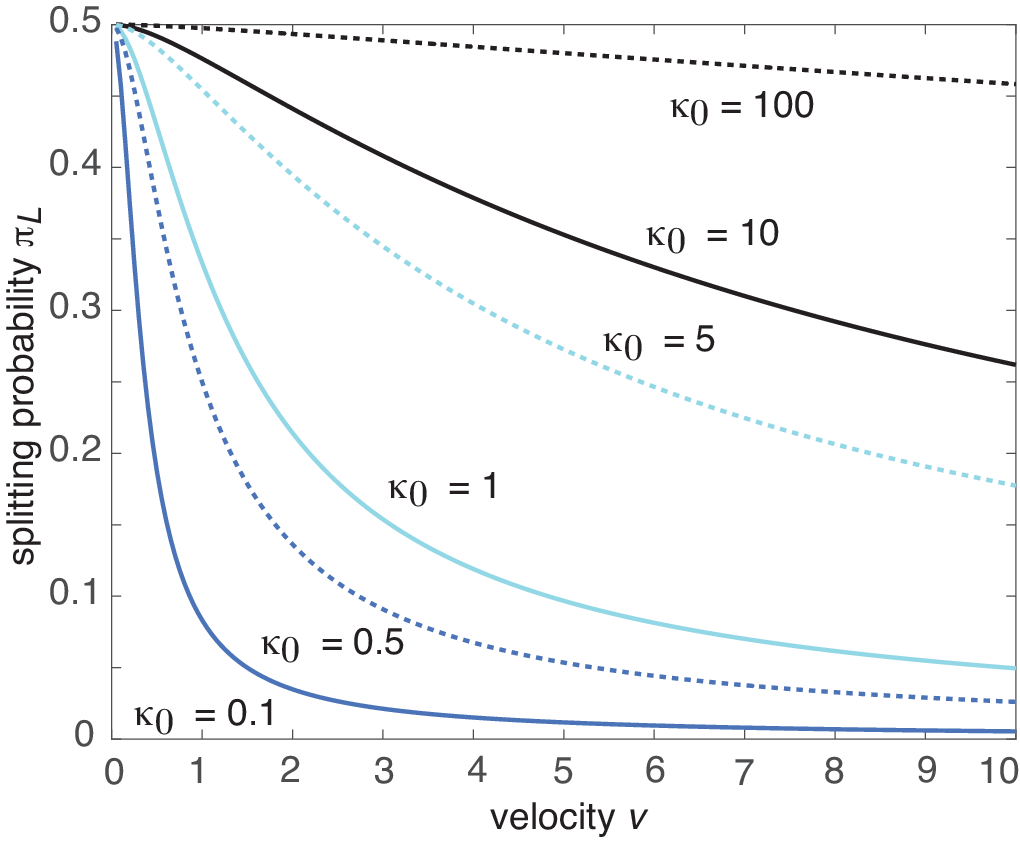}
  \caption{RTP in an interval with a partially absorbing boundary at $x=L$ and a totally absorbing boundary at $x=0$. Plot of the splitting probability $\pi_L(x_0)$ as a function of the velocity $v$ for the geometric distribution $\psi(\ell) =(1-\lambda_0)\lambda_0^{\ell}$ and various values of the absorption rate $\kappa_0$ with $\lambda_0=1/(1+\kappa_0/v)$ and $\E[\ell]/v=1/\kappa_0$. Other parameters are $L=1$, $\alpha=1$ and $x_0=0.5$.}
  \label{fig3}
  \end{figure}

 \begin{figure}[t!]
  \raggedleft
   \includegraphics[width=8cm]{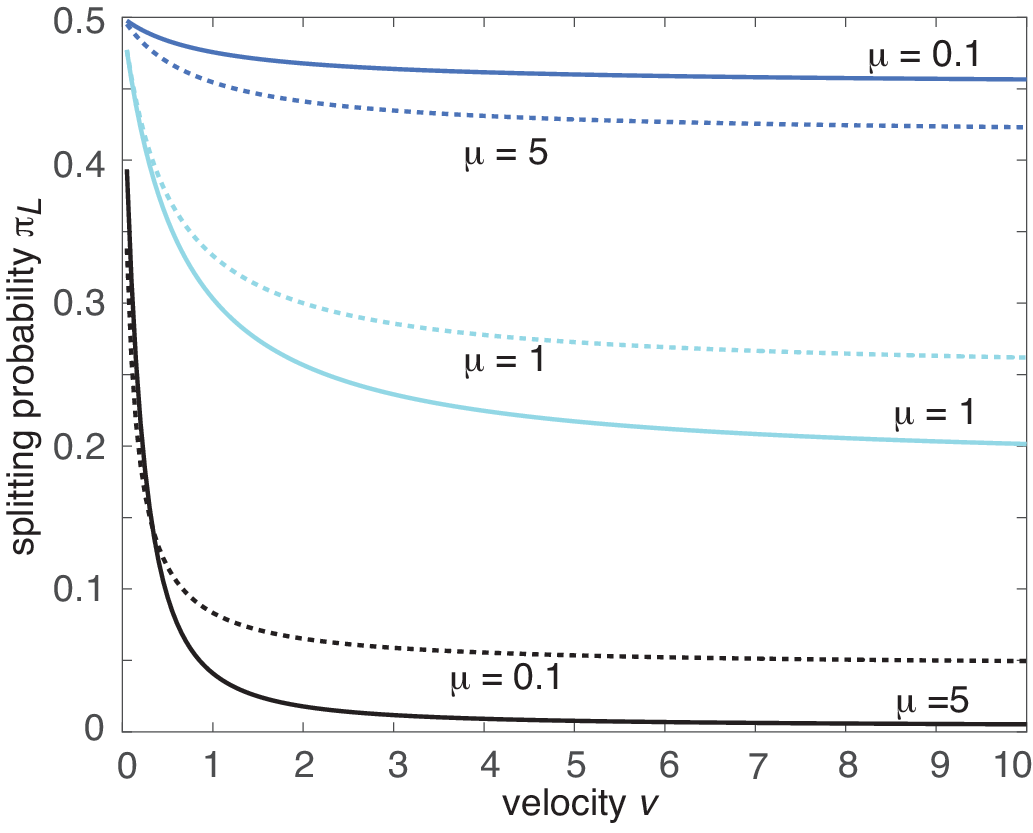}
  \caption{RTP in an interval with a partially absorbing boundary at $x=L$ and a totally absorbing boundary at $x=0$. Plot of the splitting probability $\pi_L(x_0)$ as a function of the velocity $v$ for the geometric distribution (dashed curves) and the Poisson distribution (solid curves) and various means $\mu =\E[\ell]$. Other parameters are $L=1$, $\alpha=1$ and $x_0=0.5$.}
  \label{fig4}
  \end{figure}

In Fig. \ref{fig3} we plot $\pi_L(x_0)=1-\pi_0(x_0)$ as a function of the velocity $v$ for the geometric distribution (\ref{Geom}) and various values of the absorption rate $\kappa_0$ such that $\lambda_0=1/(1+\kappa_0/v)$. (Also recall that $D_0=v^2/2\alpha$). It can be seen that $\pi_L$ is a monotonically decreasing function of $v$ for fixed $\kappa_0$. As expected, $\pi_L$ increases with the rate of absorption $\kappa_0$ such that $\pi_L\rightarrow 0$ as $\kappa_0\rightarrow 0$ and $\pi_L\rightarrow 0.5$ as $\kappa_0\rightarrow \infty$. In the case of the Poisson distribution (\ref{Poiss}) we cannot relate the distribution parameter $\lambda_0$ to an absorption rate. Therefore, in order to compare with the geometric distribution we use the same mean $\mu=\E[\ell]$ in both cases.
The results are shown in Fig. \ref{fig4}. As in Fig. \ref{fig3}, the splitting probability is a decreasing function of the velocity $v$ for fixed $\mu$. However, the dependence on the mean $\mu$ is the opposite for a Poisson distribution, namely, increasing $\mu$ decreases $\pi_L(\x_0)$. This can be understood from equations (\ref{Geom}) and (\ref{Poiss}). Since $z\in [0,1]$, it follows that $\widetilde{\psi}(z) \in [1-\lambda_0,1]$ for the geometric distribution and $\widetilde{\psi}(z) \in [\e^{-\lambda_0},1]$ for the Poisson distribution. Hence, the range of both distributions decreases as $\lambda_0$ increases. However, $\lambda_0=1/(1+\mu)$ (decreasing function of $\mu$) for the geometric distribution, whereas $\lambda_0=\mu$ (increasing function of $\mu$) for the Poisson distribution. Finally, we note that $\pi_L(x_0)$ is proportional to $ \widetilde{\psi}(\Lambda_0)$ and thus inherits its dependence on $\mu$.

\section{Conclusion}

\begin{figure}[t!]
  \centering
  \begin{tikzcd}[row sep=huge]
&\mbox{BVP for $p(\x,t|\x_0)$ and a constant reactivity z} \arrow[d] \\ &\PP(\x,z,t|\x_0) \arrow[d,"{\mathcal L}_{\ell}^{-1}"] \\ &P(\x,\ell,t|\x_0) \arrow [r,"\Psi"]  & p^{\Psi}(\x,t|\x_0)
  \end{tikzcd}
  \caption{Diagram illustrating the general steps in the encounter-based model of absorption for Brownian particles and RTPs. Let $\X(t) \in \Omega\subset \R^d$ denote the particle position and suppose that $\partial \Omega$ is a partially absorbing surface. First solve the classical BVP for the marginal density $p(\x,t|\x_0)$ in the case of a constant rate of absorption $z$. (Note that $p$ could be vector-valued in the case of a discrete set of internal states such as velocity states.) The solution of the BVP is then identified as the $z$-Laplace transform of the propagator $P(\x,\ell,t|\x_0)$ with respect to $\ell$, where $\ell(t)$ is the local time that characterizes the time that a particle is in contact with the reactive substrate over the interval $[0,t]$. The general marginal density $p^{\Psi}(\x,t|\x_0)$ is obtained by equating an absorption event with $\ell(t)$ crossing a random threshold $\widehat{\ell}$ with $\P[\widehat{\ell}>\ell]=\Psi(\ell)$. }
  \label{fig5}
  \end{figure}
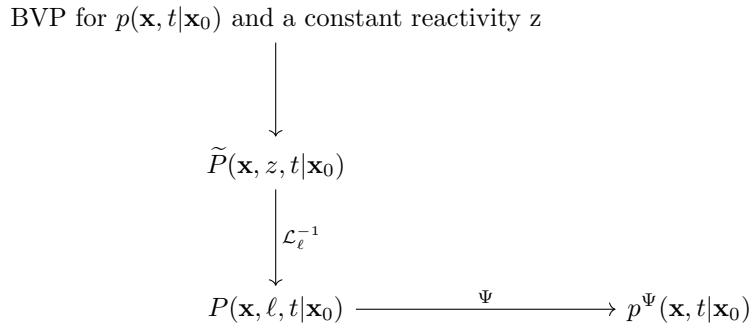

The main conclusion of the paper is that it is possible to extend the encounter-based framework for modeling Brownian motion in partially reactive media to the case of an RTP.  The main difference from the diffusive case is that the local time is now a discrete rather than a continuous random variable, which counts the number of collisions of the RTP with a totally reflecting boundary. Partial absorption is incorporated by introducing a stopping time that determines when the local time crosses a random threshold. The steps of the encounter-based method for both a Brownian particle and an RTP then proceed as follows: 
\begin{enumerate}
\item Solve the BVP for the marginal probability density $p(\x,t)$ in the case of a constant rate of absorption $\kappa_0$ using a radiation boundary condition. (This is equivalent to the Robin boundary condition in the case of diffusion.)
\item Identify $p(\x,t)$ as the (discrete or continuous) Laplace transform of the local time propagator $P(\x,\ell,t)$, in which the Laplace variable is determined by $\kappa_0$. (Both $p$ and $P$ could be vector-valued when there are internal particle states such as the velocity states of an RTP.) 
\item Invert the Laplace transform to obtain $P(\x,\ell,t)$.
\item Define the general marginal density acording to 
\begin{eqnarray*}
p^{\Psi}(\x,t)&=\int_0^{\infty}\Psi(\ell)P(\x,\ell,t)d\ell \mbox{ (Brownian particle)},\\
 p^{\Psi}(\x,t)&=\sum_{\ell =0}^{\infty} \Psi(\ell)P(\x,\ell,t) \mbox{ (RTP)}, 
\end{eqnarray*}
where $\Psi(\ell)=\P[\widehat{\ell}>\ell]$ and $\widehat{\ell}$ is the random threshold that determines when absorption occurs.
\end{enumerate}
 The general probabilistic framework is summarized in Fig. \ref{fig5}.

This work complements previous studies where we extended the encounter-based model of diffusion-mediated surface reactions to partially absorbing interior substrates \cite{Bressloff22,Bressloff22a}. For this class of problem, the particle freely enters and exits a substrate ${\mathcal U}$ and can only be absorbed when it is in the interior of ${\mathcal U}$. The basic steps of Fig. \ref{fig5} still hold, except that $\ell_t$ is now a Brownian functional known as the occupation or residence time, which specifies the amount of time the particle spends within ${\mathcal U}$ over the time interval $[0,t]$.
With regards the specific application to RTPs, there are a number of possible future directions in addition to considering partially absorbing interiors. One obvious example is a higher-dimensional model in which the RTP switches between velocity states in a continuum of different directions \cite{Bressloff11,Mori20,Santra20}. Another example is the inclusion of stochastic resetting. One useful feature of the majority of resetting protocols is that all memory of previous states of the particle are lost following reset. This leads to considerable simplification of the analysis due to the applicability of renewal theory. In the case of a standard RTP particle with reset, it is necessary to specify rules for resetting both the position and velocity state of the particle \cite{Evans18,Bressloff20,Santra20a}. As we have recently shown elsewhere for encounter-based models of diffusion \cite{Bressloff22b}, generalized models of absorption involve a memory of the contact time between the particle and reactive surface, and this must also be reset in order to exploit renewal theory. Finally, we expect the encounter-based scheme to be applicable to any stochastic search process that involves some reaction-based FPT and an appropriately defined substrate encounter time.

\end{document}